\documentclass[12pt]{article}

\usepackage{amsmath}
\usepackage{epsf}
\usepackage[dvips]{graphicx}

\begin{document}

\begin{center}
{\bfseries DIELECTRON PRODUCTION IN PP AND DP COLLISIONS} \\
{\bfseries AT 1.25 GeV/u WITH HADES}

\vskip 5mm

K.~Lapidus$^{10 \dag}$,
G.~Agakishiev$^{8}$,
C.~Agodi$^{1}$,
A.~Balanda$^{3,e}$,
G.~Bellia$^{1,a}$,
D.~Belver$^{15}$,
A.V.~Belyaev$^{6}$,
A.~Blanco$^{2}$,
M.~B\"{o}hmer$^{11}$,
J.~L.~Boyard$^{13}$
% ...

\vskip 5mm

{\small
(1) {\it
Istituto Nazionale di Fisica Nucleare - Laboratori Nazionali del Sud, 95125~Catania, Italy
}
\\
(2) {\it
LIP-Laborat\'{o}rio de Instrumenta\c{c}\~{a}o e F\'{\i}sica Experimental de Part\'{\i}culas, 3004-516~Coimbra, Portugal
}
\\
(3) {\it
Smoluchowski Institute of Physics, Jagiellonian University of Cracow, 30-059~Krak\'{o}w, Poland
}
\\
(4) {\it
GSI Helmholtzzentrum f\"{u}r Schwerionenforschung GmbH, 64291~Darmstadt, Germany
}
\\
(5) {\it
Institut f\"{u}r Strahlenphysik, Forschungszentrum Dresden-Rossendorf, 01314~Dresden, Germany
}
\\
(6) {\it
Joint Institute of Nuclear Research, 141980~Dubna, Russia
}
\\
(7) {\it
Institut f\"{u}r Kernphysik, Goethe-Universit\"{a}t, 60438~Frankfurt, Germany
}
\\
(8) {\it
II.Physikalisches Institut, Justus Liebig Universit\"{a}t Giessen, 35392~Giessen, Germany
}
\\
(9) {\it
Istituto Nazionale di Fisica Nucleare, Sezione di Milano, 20133~Milano, Italy
}
\\
(10) {\it
Institute for Nuclear Research, Russian Academy of Science, 117312~Moscow, Russia
}
\\
(11) {\it
Physik Department E12, Technische Universit\"{a}t M\"{u}nchen, 85748~M\"{u}nchen, Germany
}
\\
(12) {\it
Department of Physics, University of Cyprus, 1678~Nicosia, Cyprus
}
\\
(13) {\it
Institut de Physique Nucl\'{e}aire (UMR 8608), CNRS/IN2P3 - Universit\'{e} Paris Sud, F-91406~Orsay Cedex, France
}
\\
(14) {\it
Nuclear Physics Institute, Academy of Sciences of Czech Republic, 25068~Rez, Czech Republic
}
\\
(15) {\it
Departamento de F\'{\i}sica de Part\'{\i}culas, University of Santiago de Compostela, 15782~Santiago de Compostela, Spain
}
\\
(16) {\it
Instituto de F\'{\i}sica Corpuscular, Universidad de Valencia-CSIC, 46971~Valencia, Spain
}
\\
(a) {\it
Also at Dipartimento di Fisica e Astronomia, Universit\'{a} di Catania, 95125~Catania, Italy
}
\\
(b) {\it
Also at ISEC Coimbra, ~Coimbra, Portugal
}
\\
(c) {\it
Also at Technische Universit\"{a}t Dresden, 01062~Dresden, Germany
}
\\
(d) {\it
Also at Dipartimento di Fisica, Universit\'{a} di Milano, 20133~Milano, Italy
}
\\
(e) {\it
Also at Panstwowa Wyzsza Szkola Zawodowa, 33-300~Nowy Sacz, Poland
}
\vskip 5mm
$\dag$ {\it
E-mail: lapidus@inr.ru
}}
\end{center}


\begin{center}
\begin{minipage}{150mm}
\centerline{\bf Abstract}
Inclusive production of $e^{+}e^{-}$-pairs in $pp$ and $dp$ collisions at a kinetic beam energy of 1.25~GeV/u 
has been studied with the HADES spectrometer.
In the latter case, the main goal was to obtain data on pair emission in quasi-free 
$np$ collisions. To select this particular reaction channel the HADES experimental setup was 
extended with a Forward Wall hodoscope, which allowed to register spectator protons. 
Here, the measured invariant mass distributions demonstrate a strong 
enhancement of the pair yield for $M>140$~MeV/$c^{2}$ in comparison to $pp$ data.
\end{minipage}
\end{center}

\vskip 10mm

\section{Introduction}

HADES ({\bf H}igh {\bf A}cceptance {\bf D}i-{\bf E}lectron {\bf S}pectrometer) is located at GSI, Darmstadt and currently operated
at the SIS18 synchrotron at beam energies of 1--2~GeV/u. It is a magnetic spectrometer which is capable to register 
$e^{+}$/$e^{-}$ particles in a polar angle ranges 
from $18^{\circ}$ up to $88^{\circ}$ and has almost full azimuthal coverage. The broad experimental program includes 
the study of pair production in nucleus-nucleus collisions, elementary reactions ($pp$, $np$, $\pi p$) as well as $pA$, $\pi A$ collisions 
with the emphasis on properties of vector mesons at finite baryonic densities.

Recently, HADES has performed the study of dielectron production in $^{12}$C+$^{12}$C collisions at beam energies
of 1 and 2~GeV/u. The most important result of these studies was the
identification of a dilepton excess above expectations based 
on meson ($\pi^{0}$, $\eta$, $\omega$) decays after the chemical freeze-out \cite{Agakichiev:2006tg, Agakishiev:2007ts}. 
In the invariant mass range of $0.15<M<0.50$~GeV/$c^{2}$ and the energy of 1~GeV/u the ratio of this excess compared to the
contribution coming from $\eta$ Dalitz is found to be $F=6.8$. 
Remarkably, this result is in good agreement with previous DLS data \cite{Porter:1997rc}.

A question which arises in this context is whether the observed
discrepancy between the measured yield and 
contributions from decays of long lived resonances is due to some unique effect of nucleus-nucleus collisions 
or incomplete knowledge of dielectron production processes in elementary nucleon-nucleon interactions
at this energy regime, in particular the $np$ channel. 
Here, the bremsstrahlung process in the $np$ system is of special interest as
it was argued to be an
important source of lepton pairs;
recent consideration of this process within an One-Boson Exchange framework \cite{Kaptari:2005qz} 
predicts cross section values significantly larger than in a number of previous calculations \cite{Schafer:1989dm, Shyam:2003cn}.
However, a final consensus has not been achieved so far, 
debates concerning the magnitude of the $np$ bremsstrahlung contribution are still ongoing \cite{Shyam:2008rx}.

It is worth to mention that recent transport calculations with cross sections
for the bremsstrahlung process tuned to reproduce 
the predictions of \cite{Kaptari:2005qz} demonstrate good agreement with the DLS and HADES results \cite{Bratkovskaya:2007jk}.

This situation clearly calls for experimental data on dielectron production in elementary $NN$
collisions in the energy regime of $\sim$1~GeV. A comparison of $pp$ and $np$ channels is of particular interest.

\section{$\boldsymbol{pp}$ and $\boldsymbol{dp}$ experiments}
In 2006, the HADES collaboration has performed the study of dielectron
production in proton-proton collisions at a kinetic beam
energy of 1.25 GeV \cite{Galatyuk:2008}. Since the selected beam energy is below $\eta$ production threshold (1.27~GeV), the only sources
contributing to pair spectra are $\pi^{0}$ and $\Delta$ Dalitz decays and
small contribution from the bremsstrahlung process,
which at such energies is expected to be almost negligible as compared to $np$ case on the basis of radiation multipolarity arguments.

\begin{figure}[h]
 \centerline{
 \includegraphics[width=80mm]{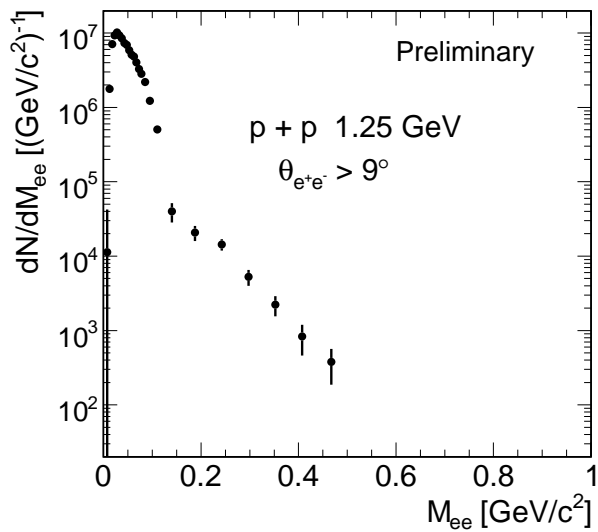}} 
 \caption{Dielectron invariant mass distribution measured in the $pp$ reaction at 1.25~GeV} \label{pp_corr}
\end{figure}

The invariant mass distribution of dielectrons measured in $pp$ collisions is shown in Fig.~\ref{pp_corr}. 
A selection on pair opening angles larger then $9^{\circ}$ was applied in order to suppress contributions from conversion photons and 
$\pi^{0}$ Dalitz decays. The combinatorial background was reconstructed by means of like-sign pair measurements and 
subtracted from the raw spectrum. The presented spectrum (and all spectra in this letter) is corrected for the
detection efficiency, however, absolute 
normalization has not yet been made.

In 2007, HADES continued its studies on $NN$ reactions with an experiment using deuteron-proton collisions.
The main goal of this experimental run was to measure pair production
properties in the $np$ channel at the same beam energy per nucleon as in the $pp$ case. 

Within impulse approximation (IA) \cite{Chew:1952fb} the collision of a high
energy deuteron with a nucleon is reduced to the quasi-free
nucleon-nucleon interaction: one of the nucleons forming the deuteron does not participate in the reaction and acts
as a spectator. Since typical values of Fermi momenta are small compared
to those of the beam, 
the spectator nucleon carry approximately half of deuteron momentum and moves
at small polar angle in the laboratory frame.
These features allows to tag the quasi-free $np$ channel by requiring a spectator proton at very forward directions. 
In order to be capable of such technique the HADES setup was upgraded with a Forward Wall (FW) scintillator hodoscope.

The FW is an array which consists of nearly 300 scintillating cells with each 2.54 cm thickness. 
During the $dp$ experiment it was located 7 meters downstream the target and covered polar angles from $0.33^{\circ}$ up to $7.17^{\circ}$.
Our Monte Carlo studies show that about $90\%$ of all spectator protons in the
reaction under consideration are inside the FW acceptance.

Our main experimental trigger used 
during the $dp$ experiment was configured in such a way that 
the dielectron pair was detected in coincidence 
with a charged particle in the FW. 
Time of flight and coordinate measurements with the FW detector allowed us to 
reject spurious signals (e.g. coming from photons or $e^{+}$/$e^{-}$) and to
suppress contributions of quasi-free 
$pp$ scattering by selecting particles having kinematical properties of the spectator.

\begin{figure}[h]
 \centerline{
 \includegraphics[width=140mm]{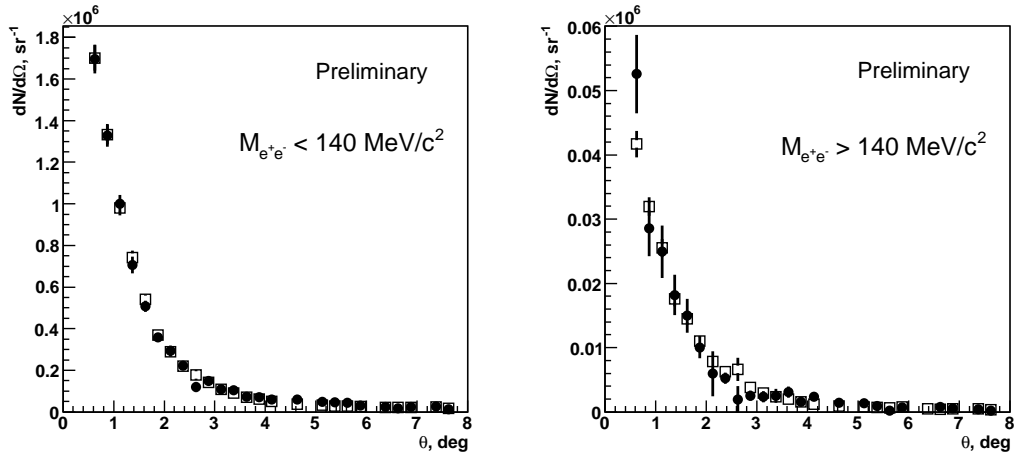}}
 \caption{Angular distribution of particles in the FW hodoscope, in coincidence with an $e^{+}e^{-}$-pair.
          Left: for invariant mass $M<140$~MeV/$c^{2}$, right:
          $M>140$~MeV/$c^{2}$. The full circles represent experimental data,
          whereas the open squares
          show predictions of a Pluto simulation} \label{FW_theta_lab_sim}
\end{figure}

Fig.~\ref{FW_theta_lab_sim} (full circles) present the polar angle
distributions in the laboratory frame of charged particles
in coincidence with \mbox{$e^{+}e^{-}$-pair} production and detected in the FW.
Two regions of pair invariant mass are examined separately: $M<140$~MeV/$c^{2}$ (left) and $M>140$~MeV/$c^{2}$ (right).
As one can see, the angular distributions are 
peaked at very small angles and rapidly drops with growing angle.
Such a form of distribution is in qualitative agreement with IA picture previously outlined.
Moreover, the distributions associated with two different mass components of dielectron spectra are
essentially the same. This observation serves as a basic test of the IA applicability
for a treatment of $dp$ reaction in our conditions.
The open squares on Fig.~\ref{FW_theta_lab_sim} demonstrate predictions
of the Monte Carlo event generator Pluto \cite{Fro:2007} which relies on the IA for the simulation of the $dp$ reaction.
A detailed description of the Pluto simulations for $pp$ and $dp$ reactions will be given in a forthcoming publication.
The results of simulations were scaled to the same integral values as experimental data. As follows from Fig.~\ref{FW_theta_lab_sim}
the Pluto simulations describe the polar angle distribution shapes in both mass regions very well.

\begin{figure}[h]
 \centerline{
 \includegraphics[width=80mm,height=80mm]{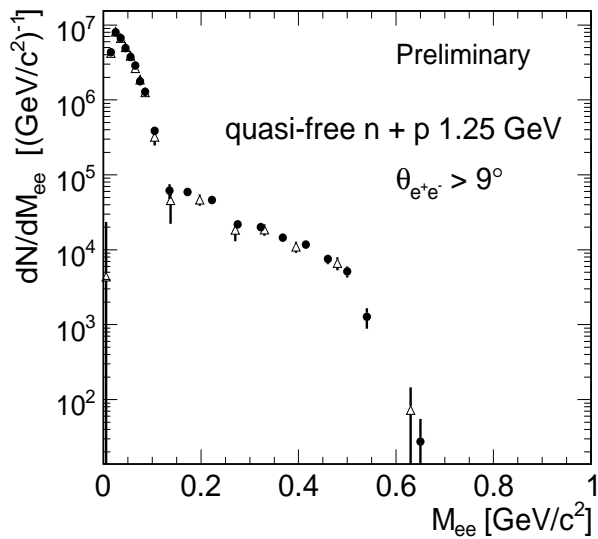}}
 \caption{Dielectron invariant mass distribution measured in the quasi-free $np$ reaction at 1.25~GeV. 
          The full circles corresponds to spectator tagging within full FW
          acceptance, whereas the open triangles are obtained with a selection
          on a more
          restricted polar angle range of $\theta_{FW}<1^{\circ}$} \label{ishepp_np_corr_vs_theta_less_1}
\end{figure}

The invariant mass distribution of $e^{+}e^{-}$-pairs collected in the $dp$ experiment 
with a selection on the quasi-free $np$ component is 
presented in Fig.~\ref{ishepp_np_corr_vs_theta_less_1} (full circles).
Apparently there is a drastic difference between $pp$ and $np$ spectra measured with HADES.
The pair invariant mass spectra in $np$ spans almost to 700~MeV/$c^{2}$ and has an enriched
pair yield in the region of invariant masses $M > 140$~MeV$/c^{2}$ as compared
to the $pp$ data. 
Because of Fermi motion inside deuteron, the energy accessible in quasi-free $np$ reactions
is larger in comparison with $pp$ collisions. Thus part of the observed
difference in pair yield above $\pi^{0}$ Dalitz region is likely to be caused
by subthreshold processes 
(production of $\eta$ meson and, possibly, baryonic resonances above $\Delta$). 
The ongoing analysis and 
comparison with simulation is aimed 
to separate effects connected with Fermi motion from pure isospin origin.

Besides providing experimental trigger conditions 
on quasi-free $np$ interactions, the FW allows to perform more 
detailed investigation of this particular reaction by studying characteristics of pair production as
a functions of FW observables (angular/momentum measurements, multiplicity, etc).
One of such applications is to study the dependence of the invariant mass spectrum 
as a function of the polar angle of the detected spectator proton. 

Indeed, at rather large polar angle values (but still in FW acceptance) one can speculate
that a number of effects (e.g. final state interaction, violation of IA) can occur
and affect the observed dielectron invariant mass distribution. 
A prompt method to investigate/reject effects of such a kind is to restrict measurements 
to very small values of spectator polar angle, where it has very small values of transverse momentum, 
thus ensuring a safe regime of spectator tagging.

The open triangles on the Fig.~\ref{ishepp_np_corr_vs_theta_less_1} 
represent the $e^{+}e^{-}$-pair invariant mass distribution 
associated with spectator protons detected in polar angle less than $1^{\circ}$. 
This spectrum was normalized to the same $\pi^{0}$ yield as in the total spectrum (full circles)
in order to facilitate the comparison of the shape of 
the massive components of both spectra. It is clearly seen that
in the region $M>140$~MeV/$c^{2}$ the two spectra completely overlap, meaning that 
the observed intense emission of massive dielectrons does not correlate with the polar angle of spectator proton.
Studies of such
kind allow to put stricter constraints 
on theoretical predictions for pair production in $np$ collisions.

\section{Summary}

In summary, we presented preliminary results on 
inclusive $e^{+}e^{-}$-pair production studied in $pp$
and $dp$ collisions at beam energies of 1.25 GeV per projectile nucleon with the
HADES spectrometer.
In order to investigate the quasi-free $np$ scattering, the HADES setup was equipped
with a Forward Wall hodoscope, which provided the possibility to detect 
spectator protons at small polar angles. 
Spectra measured in quasi-free $np$ collisions demonstrate a strong enhancement 
of the pair yield above 140~MeV/$c^{2}$ as compared to the $pp$ case. 
Undoubtedly, the investigation of electron-positron pair production in elementary collisions 
is a significant step in revealing the nature of excess observed in nucleus-nucleus collisions.
More detailed information will be obtained through a comparison of measured spectra with theoretical predictions.

\section*{Acknowledgments}
\renewcommand{\baselinestretch}{0.9}
\footnotesize
The HADES collaboration gratefully acknowledges the support
by BMBF grants 06MT238, 06TM970I, 06GI146I,
06F-140, 06FY171, and 06DR135, by DFG EClust 153
(Germany), by GSI (TM-KRUE, TM-FR1, GI/ME3, OF/STR),
by grants GA AS CR IAA100480803 and MSMT LC 07050 (Czech Republic), 
by grant KBN 5P03B 140 20 (Poland),
by INFN (Italy), by CNRS/IN2P3 (France), by grants
MCYT FPA2006-09154, XUGA PGID IT06PXIC296091PM
and CPAN CSD2007-00042 (Spain), by grant FTC POCI/FP
/81982 /2007 (Portugal), by grant UCY-10.3.11.12 (Cyprus),
by INTAS grant 06-1000012-8861 and EU contract RII3-CT-2004-506078.
\renewcommand{\baselinestretch}{1}

\end{document}